\documentclass[
    ,final            
  ]
  {aipproc}

\layoutstyle{8x11double}

\begin{document}

\newcommand\rxte{\textsl{RXTE}}
\newcommand\pca{\textsl{PCA}}
\newcommand\hexte{\textsl{HEXTE}}
\newcommand\xmm{\textsl{XMM-Newton}}
\newcommand\chandra{\textsl{Chandra}}
\newcommand\integral{\textsl{INTEGRAL}}
\newcommand\xeus{\textsl{XEUS}}
\newcommand\maxim{\textsl{MAXIM}}
\newcommand\conx{\textsl{Constellation-X}}
\newcommand\astro{\textsl{Astro E II}}
\newcommand\swift{\textsl{Swift}}
\newcommand\asm{\textsl{ASM}}

\title{The Future of X-ray Spectroscopy of \\ Galactic Black Hole Binaries}

\author{Michael A. Nowak}{
  address={MIT-CXC, NE80-6077, 77 Massachusetts Ave., Cambridge, MA
02139; mnowak@space.mit.edu}
}

\begin{abstract}
There are four major X-ray satellites currently in operation (\rxte,
\chandra, \xmm, \integral), with two more shortly to follow (\astro,
\swift), and several very ambitious observatories in various stages of
planning (\conx, \maxim, \xeus).  This very rich period of X-ray
observation is leading to great advances in our understanding of the
accretion flow onto the black hole, although we are quickly learning
(or perhaps better put, remembering) exactly how complicated this
flow can be.  This review was meant to assess future
prospects for X-ray spectroscopy of black hole binaries; however, I first look
backward to the observations and theories that helped us arrive at
our current `paradigm'.  I then discuss current and near-future spectroscopic
studies, which increasingly (and very fruitfully) treat X-ray
spectroscopy as part of a larger, intimately connected picture along
with radio, optical, and gamma-ray spectroscopy.  Equally importantly,
and in large part thanks to the success of \rxte, there is now a
strong realization that spectral-temporal correlations, even across
wavelength bands, are crucial to our understanding of the physics of
these systems.  Going forward, we are well-poised to continue to
advance our knowledge via X-ray spectroscopy, both with existing
satellites that have a long lifetime ahead of them (\chandra, \xmm,
\integral), and with the next generation of instruments.  If there is
any `hole' in this bright future, it is the potential loss of
\rxte, with no designated follow-up mission.  Studies of
multi-wavelength spectral-temporal correlations will become more
difficult due to the loss of two important attributes of \rxte: its
fast timing capabilities and its extremely flexible scheduling which
has made many of these studies possible.

\end{abstract}

\maketitle

\section{Looking Backward}

\textsl{Those who forget history are condemned to repeat it. --
George Santayana}
\smallskip

\begin{figure}
  \includegraphics[width=.4\textwidth]{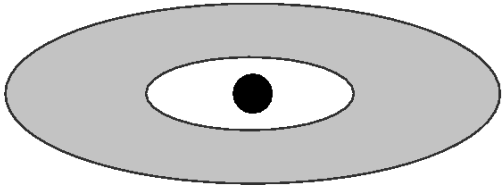}
  \caption{The spectroscopic model of accretion flows in black hole
  binaries, circa 1973 (e.g., Shakura \& Sunyaev 1973).}\label{fig:alpha}
\end{figure}

There are a number of reviews describing theoretical models and
observations of galactic black hole candidate (GBHC) binaries
\cite[e.g.,][]{done:01b,reynolds:03a}. It is interesting to look back,
however, and note that many of the components incorporated into models
today had their origins some time ago, with many important insights
made using very sparse data. Some ideas have come into and out of
consideration several times over the past thirty years.  See
\cite{reynolds:03a}, for example, for a brief description of the history of the
fluorescent Fe line in GBHC, which has gone from being interpreted as
broad to narrow, and back and forth again, several times
\cite[][etc.]{barr:85a,fabian:89a,kitamoto:90a,done:92a}. I myself have
been on both sides of this issue \cite{wilms:99aa,nowak:02a}, although
I am certainly not alone in this regard \cite{done:92a,done:99a}.

The seminal theoretical work that helped usher in the ``modern era''
of the study of accretion flows was that of Shakura \& Sunyaev
\cite[][see Fig.~\ref{fig:alpha}]{shakura:73a}. Theories, aided by
observations, quickly added complexity to this basic picture.  The
concept of a two-phase flow (i.e., disk and `corona') was introduced,
and was even suggested to represent `advection domination'
\cite{shapiro:76a,ichimaru:77a}.  Disks were hypothesized to be warped
\cite{petterson:77a}, to produce magnetic flares in their inner regions
\cite{galeev:79a}, or to be surrounded by a hot corona that produced 
the characteristic spectrum of `hard state' observations
\cite{sunyaev:79a}.  Very brief, fast-photometry optical observations
\cite{motch:83a} revived the concept of disk flares, and tied them to
optical synchrotron emission
\cite{fabian:82a}.  Magneto-hydrodynamic turbulence was suggested as
the viscous dissipation mechanism \cite{balbus:91a}, and later even
suggested to be acting within the innermost stable circular orbit
(ISCO) \cite{agol:01a}.

A major observational advance was made when radio jets were discovered
in X-ray binaries \cite{mirabel:94a}.  It is interesting to note,
however, that `radio jet ejection events' previously had been
hypothesized, probably correctly so, based solely upon `dipping
events' in X-ray observations of GX~339$-$4 \cite{miyamoto:91b}.
Meanwhile, the study of warped disks had been revived
\cite{pringle:96a}, and advection dominated flows had undergone
several revivals \cite{rees:82a,narayan:95a} and further had winds,
and several new acronyms, appended to them \cite{blandford:99a}.  In
at least one instance \cite{corbel:02b}, an X-ray binary jet clear
manifested itself in the X-ray due to its interaction with the
interstellar medium. It also has been suggested, however, that steady
jets in the hard state can significantly contribute to the observed
2-200\,keV X-ray spectrum \cite{markoff:01a}.

\begin{figure}
  \includegraphics[width=.4\textwidth]{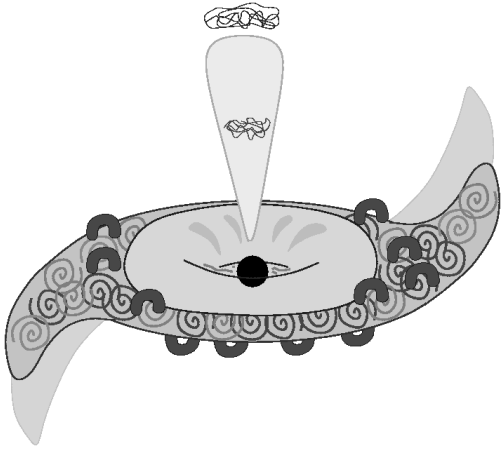} \caption{The
  spectroscopic model of accretion flows in black hole binaries, circa
  2003.  It is now widely believed that (starting from the outer disk,
  and working our way inward, then along the axis and upward) that the
  accretion flow consists of: a warped disk, with disk wind; MHD
  turbulence providing the viscosity; magnetic flaring activity and/or a
  corona; dissipation from the region interior to the innermost stable
  circular orbit; an outward propagating jet yielding radio and
  possibly X-ray emission; and the interaction of this jet with the
  surrounding interstellar medium.}\label{fig:omega}
\end{figure}

This past 30 years of research has led to a picture much as presented
in Fig.~\ref{fig:omega}.  My own reading of our field is that most of
us would agree that \emph{all} the components shown in
Fig.~\ref{fig:omega} are relevant.  The major questions, then, relate
not to the existence of these components, but rather to their relative
contributions to the observed spectra (and variability) from source to
source, and, within a given source, as a function of the source
luminosity, `state', and history.  Ultimately, our goal is to tie
these phenomena to basic system parameters (black hole mass, spin,
accretion rate onto the hole, secondary mass, binary separation,
etc.), and use them to study General Relativity in the strong field
regime \cite[e.g.,][]{wilms:02a} and plasma physics under extreme
conditions.

\section{The `Golden Age'}

As discussed in the abstract above, we are perhaps living in the
``Golden Age'' of X-ray spectroscopy, with four currently operating
instruments that are in many ways complementary to one another. Using
various combinations, one can obtain spectra in the 0.1--600\,keV
regime, (continuous) timing ranging from $\mu$sec to 100's of ksec
(and 10's of Msec, if one includes the \rxte-\asm), and spectral
resolutions as large as $E/\Delta E \approx 1000$.  Coordinated
multiple X-ray satellite observations are now routinely performed. For
example, \rxte\ provides the broad band spectrum while, \chandra\
allows the Fe line region to be decomposed into broad and narrow
components \cite[e.g.,][]{miller:02a}.  

\begin{figure*}
  \includegraphics[width=0.8\textwidth]{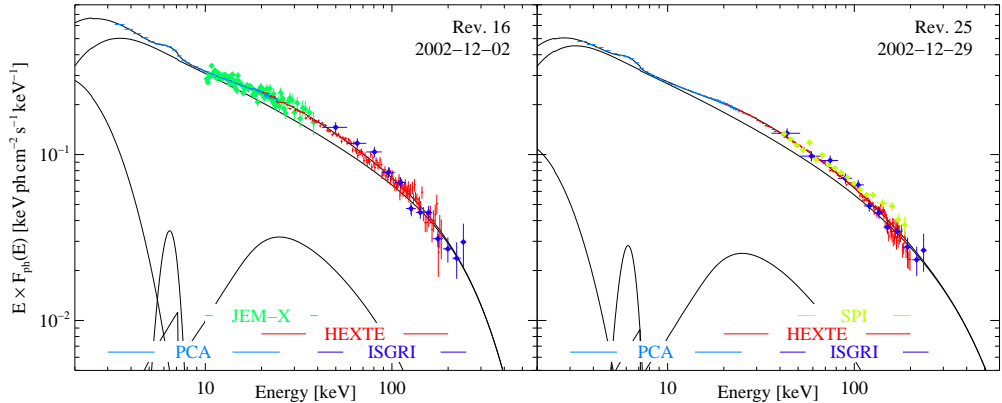}
  \caption{Simultaneous \textsl{RXTE} and \textsl{INTEGRAL}
  observations of Cyg X-1 (Pottschmidt et al. 2003). These are
  `unfolded spectra', comprised of a disk component, a broad gaussian
  line at 6.4\,keV, and a power law with exponential cut-off reflected
  from an ionized slab.  In the regions of overlap, there is very good
  agreement between the \textsl{RXTE} instruments (\textsl{PCA} and
  \textsl{HEXTE}), and the \textsl{INTEGRAL} instruments
  (\textsl{JEM-X} and \textsl{HEXTE}).}\label{fig:int}
\end{figure*}

In Fig.~\ref{fig:int}, I show joint
\rxte-\integral\ observations of Cyg~X-1 \cite{pottschmidt:03a}, 
which provides an extremely broad-band spectrum.  These particular
spectra are very well fit by a low temperature, disk blackbody,
Compton upscattered in an $\approx 100$\,keV corona with optical depth
$\tau_{\rm es} \approx 1$.  The spectra are further reflected off of a
cold, mildly ionized slab, with reflection fraction $\Omega/2\pi
\approx 0.2$.  There are plans to follow-up this particular observation in the
Fall of 2004 with a multi-observatory campaign (PI: J. Wilms) that
will consist of ground based radio and optical, along with
simultaneous \xmm, \rxte, and \integral\ observations. This will
achieve the broadest band spectrum of any GBHC to date.

This highlights a very important point about X-ray spectroscopy as
currently practiced: it no longer solely deals with X-ray
spectra. Gamma-ray spectra are crucial for constraining high energy
cutoffs, which yield coronal temperatures and are likely important
for distinguishing between X-ray emission from jets and coronae.
Radio spectra constrain models of the jets.  IR and optical spectra
constrain jet and outer-disk models.  All these components are coupled
observationally, and hence must be coupled theoretically.  One of
\rxte's greatest contributions to the study of X-ray spectra of GBHC
has been to reveal the coupling of radio and X-ray emission.  For
example, the low/hard state of GX~339$-$4 reveals that the X-ray flux,
$F_{\rm X}$, is related to the radio flux, $F_{\rm r}$, by $F_{\rm X}
\propto F_{\rm r}^{1.4}$ \cite{corbel:03a}.  It further has been
suggested that this trend may be universal in the hard state
of GBHC \cite{gallo:03a}.

Whether one agrees or disagrees with the X-ray jet model of hard state
GBHC \cite[e.g.][]{markoff:01a}, it is worthwhile noting that prior to
the launch of \rxte, few would have ever even attempted to apply a
model (at least to GBHC, as opposed to AGN) that attempts to describe
the spectrum over 9 orders of magnitude in photon energy.  The data
simply did not exist.  The extremely flexible scheduling of \rxte\ has
allowed such multi-wavelength observations to be obtained far more
readily.  Furthermore, such observations have been carried out over
multiple flux levels and spectral states\footnote{This points out
another unique advantage of \rxte. To date, there have been eight
\chandra\ observations of Cyg~X-1, but over 200 \rxte\ observations.
If a single observation allows one to study `weather', \rxte\ has
allowed us to study GBHC `climates'.}, allowing the discovery of
spectral correlations as described above \cite{corbel:03a,gallo:03a}.

Along with broad-band flux and spectral correlations, current \rxte\
observations have been highlighting correlations of these properties
with variability features. In the hard state, flux appears correlated
with spectral hardness, which in turn appears correlated with peak
frequencies of characteristic broad features in power spectra (PSD) of
X-ray variability \cite[e.g.][]{dimatteo:99a,gilfanov:99a,nowak:02a}
and with the time lags between hard and soft X-ray variability
\cite{pottschmidt:00a,nowak:02a,pottschmidt:02a}.  These properties may
further be correlated with `finer' spectral features, such as
reflection fraction \cite{zdziarski:99a,gilfanov:99a,nowak:02a}.
Again, these strong observational couplings indicate that there must
be fundamental theoretical underpinnings. X-ray spectroscopy is (or at
least, should be) inseparable from X-ray variability studies.

\section{Looking Forward}

\textsl{Greetings, my friends.  We are all interested in the
  future, for that is where you and I are going to spend the rest of
  our lives.  And remember, my friends, future events such as these
  will affect you, in the future. -- Criswell, `Plan 9 From Outer Space'}

\subsection{The Glass is (Mostly) Full}

Where do we go from here, and are we well-prepared to get there? In
the short-term, over the next two years, I believe that the answer is
an emphatic yes.  We will not see any of our capabilities diminish
(barring any unforeseen events that affect currently operating
satellites), and we will see some very important new capabilities
emerge.  One of these capabilities that I personally am most excited
about is the advent of new IR and optical observations of GBHC.  Although the
discovery of microquasars is nearly a decade old, dedicated
radio/X-ray spectral campaigns are less than seven years old.  The
correlation of these properties with X-ray timing properties is even
more recent.  Extending such studies to the IR and optical regimes is
likely to be very important.

I foresee this progressing in two ways, both of which are underway now.
First, there are dedicated, frequent optical observations of GBHC with
small telescopes, such as recent optical observations of
XTE~J1550$-$564 \cite{jain:00a}.  Such observations can be correlated
with the daily monitoring by the \rxte-\asm, which will likely shed
further light on the nature of GBHC state transitions.  IR/optical
studies can also provide needed information about the transitions from
the radio (jet?)  to the X-ray (corona?) part of the spectrum.
Specifically, we only have limited observational knowledge of the IR
turnover between these two spectral regimes \cite[although
see][]{corbel:02a}.

Second, the twenty year old suggestion that inner-disk flares create
optical synchrotron emission \cite{fabian:82a} was based upon only 100
seconds worth of data, where it was unclear whether the optical led or
trailed, or was correlated or anti-correlated \cite{motch:83a}!
Observations with large modern telescopes with fast photometry
systems, such as the \textsl{VLT}, have dramatically improved this
situation.  Very exciting examples are the recent optical/X-ray
observations of XTE~J1118$+$480, where possibly correlated optical and
X-ray quasi-periodic oscillations (QPO) are observed, and where it was
further suggested that the optical variability traces a synchrotron
component of the spectrum \cite{hynes:03b}.  Compared to correlated
radio/X-ray spectral-temporal studies, these optical/X-ray studies are
only in their infancy.  Again, however, they do rely on the future
availability of a flexible, easily scheduled X-ray observatory.

\begin{figure}
  \includegraphics[width=.4\textwidth]{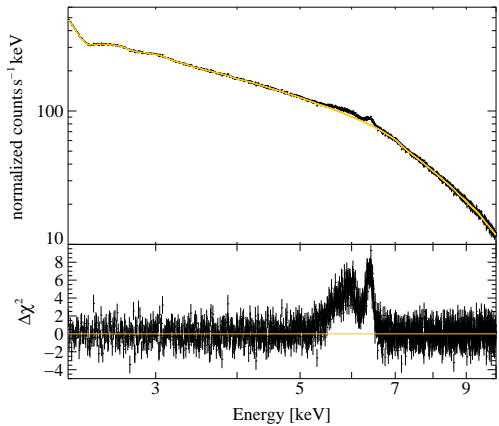} \caption{To a limited
  extent, \textsl{XMM-Newton} has the potential for replacing
  \textsl{RXTE} for studies of X-ray spectral-temporal correlations.
  Shown here is a simulation of the Cyg X-1 hard state spectrum
  observed in a proposed modification of the \textsl{XMM-Newton}
  ``timing mode'' (Wilms et al., in prep.).  The simulated spectrum is a
  reflected power law, plus broad and narrow Fe line components,
  absent from the fit, with residuals shown above.}\label{fig:xmm}
\end{figure}

Spectral-temporal X-ray studies will undoubtedly continue in the near
term, especially with \rxte.  But what about the post-\rxte\ era?
\chandra\ and the imminent \astro\ have somewhat reduced effective 
areas compared to \rxte, as well as potential problems with photon
pile-up \cite[e.g.,][]{davis:01a} for sources as bright as many
GBHC. \xmm, although having good effective area, CCD resolution, and
reasonably fast timing capability, has suffered from telemetry
constraints for bright sources\footnote{This has often required the
use of the `burst mode' for bright GBHC, wherein only $\approx 3\%$ of
the photons are telemetered to ground.}.  A potential work-around for
this situation may offer the possibility of dramatically improving the
utility of \xmm\ for GBHC spectral-temporal studies (Wilms et al., in
prep.).  In a suggested new mode, the lower energy threshold will be
raised to $\approx 2$\,keV, thereby reducing telemetry and allowing
the use of `timing mode'.  If successful, this will allow \rxte-like
timing with CCD spectral resolution (see Fig.~\ref{fig:xmm} for a
spectral simulation of Cyg X-1; courtesy J. Wilms).  However, \xmm\
still has fairly severe scheduling constraints, and such a new mode
will not be effective for GBHC much brighter than Cyg~X-1 (which
numerous X-ray novae in outburst are).

Regardless of the success of this proposed modified mode, in the near
term \xmm\ and \chandra\ both will further our understanding of the
X-ray spectra, and variability, of quiescent GBHC.  (This is provided
that these objects are studied with sufficiently long integration
times; short observations are often limited to simple spectra,
e.g. constrained power laws, and flux measurements.)  One interesting
recent GBHC observation is of the quiescent state of XTE~J1650$-$500
\cite{tomsick:03a}. The observed spectrum was hard, as is typical for
`low/hard state' GBHC.  Furthermore, the X-ray variability revealed a
break in the power spectrum at very low frequency, consistent with
previously observed trends for the characteristic PSD frequencies of
hard state GBHC to decrease with decreasing flux.  This is the first
claim that this trend continues into such extremely faint states.

\begin{figure}
  \includegraphics[width=0.4\textwidth]{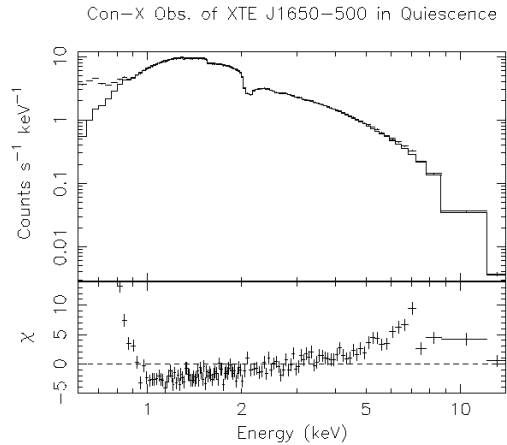} \caption{A simulated
  50\,ksec \textsl{Constellation-X} observation of a quiescent black
  hole binary, such as XTE J1650-500.  The simulated model is a disk
  component, reflected power law, and broad line, but only the power
  law has been fit.  Residuals clearly reveal the disk and broad line
  components of the spectrum.} \label{fig:conx}
\end{figure}

\begin{figure}
  \includegraphics[width=.4\textwidth]{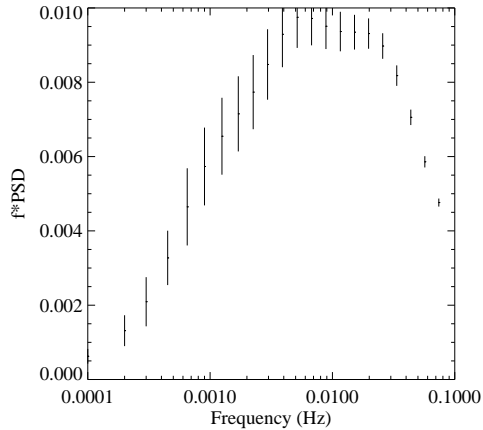} \caption{Simulated
  50\,ksec \textsl{Constellation-X} observation of the X-ray
  variability power spectrum (shown as frequency $\times$ amplitude)
  possibly associated with the quiescent X-ray spectrum shown in
  Fig.~5. (Power spectrum amplitude and frequency extrapolated from
  \textsl{Chandra} observations of XTE~J1650$-$500 in quiescence;
  Tomsick, Kalemci, \& Kaaret 2003.)}\label{fig:cont}
\end{figure}

This is one area where the farther future offers substantially greater
promise.  Current studies are limited by photon statistics of such
faint sources. (For a given signal-to-noise, the required integration
times for PSD studies scale as received count rate squared; see
\cite{vanderklis:95a}.)  An important attribute of \conx\ (as well as \xeus,
although here I do not show simulations of this latter mission) is
that with its proposed very large effective area, we simultaneously will obtain
detailed X-ray spectral and variability data from extremely faint sources. As
an example, in Fig.~\ref{fig:conx} I present a simulation of a
50\,ksec observation of a quiescent GBHC, such as XTE~J1650$-$500.
Instead of simple power-law spectrum, I have included a disk blackbody
and relativistic line, which are clearly revealed in the residual
spectra when only fitting a power law\footnote{This simulation and
figure are modeled after a similar presentation by Jon Miller at the
Constellation-X Workshop held at Columbia University, May 2003.}

In addition to revealing structure in the spectrum, \conx\
observations potentially could measure structure in the PSD of the
X-ray variability.  Fig.~\ref{fig:cont} shows a simulated PSD that may
be associated with the spectrum of Fig.~\ref{fig:conx}. (The
variability parameters are modeled after the observations presented in
\cite{tomsick:03a}.)  Thus, by correlating such spectral features as
the presence of a soft disk component or broad line with
characteristic variability features, we may determine whether these
properties truly are associated, for example, with a varying
`transition radius' between an outer thin disk and an inner corona
\cite[see the reviews of][and references therein]{done:01b,reynolds:03a}.

\subsection{The Glass is (a Little) Empty}

As sketched out above, I am very optimistic about current progress in
the field, and about some of the new directions that the field seems
to be taking.  But is there anything missing from our spectroscopic
capabilities in the future?  And is there room for a replacement for
\rxte?  My own opinion is that the answer to both questions is `yes'.
Two of my major concerns for the future have been alluded to above.
The first is that we have become quite attached to the flexibility,
ease of scheduling, and the --- still, in comparison to other X-ray
satellites --- rapid maneuverability of \rxte, yet we have no
designated replacement.

The \swift\ X-ray/gamma-ray satellite will of course be flexible,
rapidly maneuverable, and have broad energy coverage with good timing
capabilities.  Understandably, however, the majority of its program
will be devoted to its gamma-ray burst program, and we cannot
reasonably expect it to supplant the current radio/optical/X-ray
multi-wavelength spectral programs conducted by \rxte.  To a limited extent,
\swift\ also will act as a replacement for the \rxte-\asm.  Again,
however, this replacement will be incomplete.  The eventual loss of
the \asm\ will impact future \xmm\ and \chandra\ observations in
several ways.  

Not only will we no longer have a soft X-ray trigger
for scheduling pointed spectral observations of rare events, but we
will also no longer have a long term lightcurve to act as context for
pointed observations by other instruments.  As alluded to above, the
number of \xmm\ or \chandra\ observations of any given GBHC is usually
quite limited in comparison to available \rxte\ observations.
Currently, these more limited observations can be compared to spectral
properties revealed by the \asm, which in turn often can be compared
to more detailed pointed \rxte\ observations of the same source at an
earlier time with similar \asm\ spectral characteristics.  (Currently,
many radio/optical/X-ray studies are conducted as monitoring programs
utilizing the \asm, e.g., \cite{jain:00a,corbel:02a}.)  We very well
may lose this ability well before the end of the mission lifetime of
either \xmm\ or \chandra.

The second concern that I have for future X-ray spectral studies is
that we may be forgoing opportunities to delve deeper into studies of
spectral-temporal correlations.  Further progress is partly contingent
upon having the ability to study rapid variability with a large
effective area instrument.  Specifically, the ability to study X-ray
variability \emph{phase information} and \emph{coherence} is severely
limited by photon statistics, even more so than are PSD studies
\cite{vanderklis:95a,vaughan:97a}. 

One currently utilized method of studying spectral-temporal
correlations is the so-called `Fourier resolved spectroscopy'
\cite{gilfanov:99a,revnivtsev:01a} (although this technique has been
previously applied to \textsl{Ginga} observations of GBHC).
Essentially, it involves weighting a spectrum by a PSD amplitude.  One
of my major objections to such techniques is that they ignore phase
information.  For example, a pivoting power law appears as a broken
spectrum, with the break at the pivot.  Or another way of phrasing it,
we are taking knowledge from an incoherent sum (the PSD), and applying
that to something which is likely comprised of (quasi-)coherently added
components (the spectrum).  Techniques that go from the spectrum to an
incoherently summed PSD seem to me more promising.

Such thoughts, as with many of the theories discussed in the
introduction, are of course not new, and have been contemplated for
prior X-ray observations of GBHC.  For example, using \textsl{Ginga}
data, Miyamoto and collaborators hypothesized that GBHC variability
was comprised of `disk' and `coronal' components incoherently summed
\cite{miyamoto:94a}.  By first fitting the spectral component and then
using the normalizations of the disk and coronal portions of the
spectrum, they found that a fair representation of the associated PSD
could be obtained.  An incoherent sum for the variability, however, is
just the first approximation.  Each component likely has its own phase
(or, equivalently, lag between hard and soft variability) that can
add/interfere in the regions of strong overlap of the PSD components.

\begin{figure*}
  \includegraphics[height=.46\textwidth]{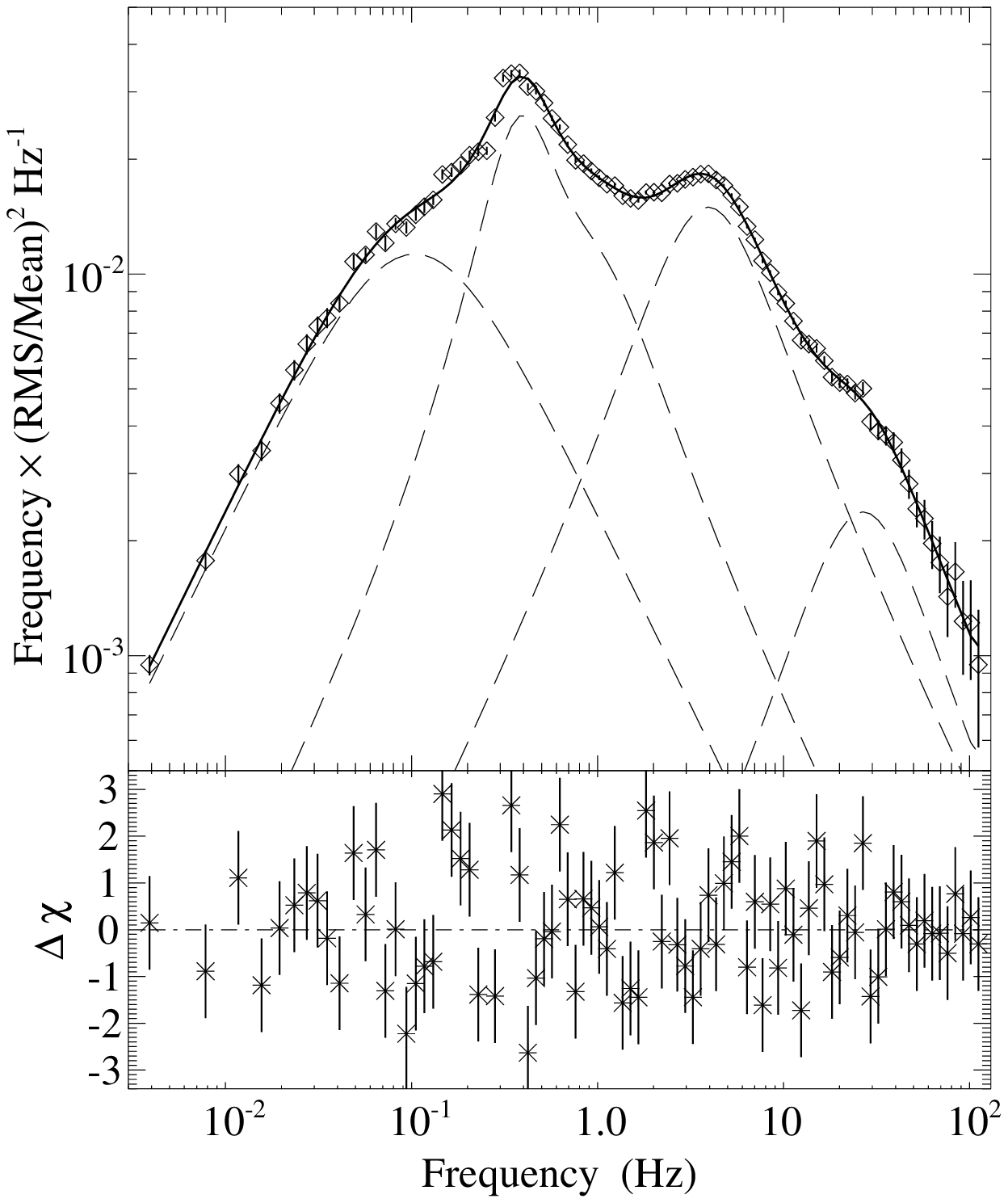}
  \includegraphics[height=.445\textwidth]{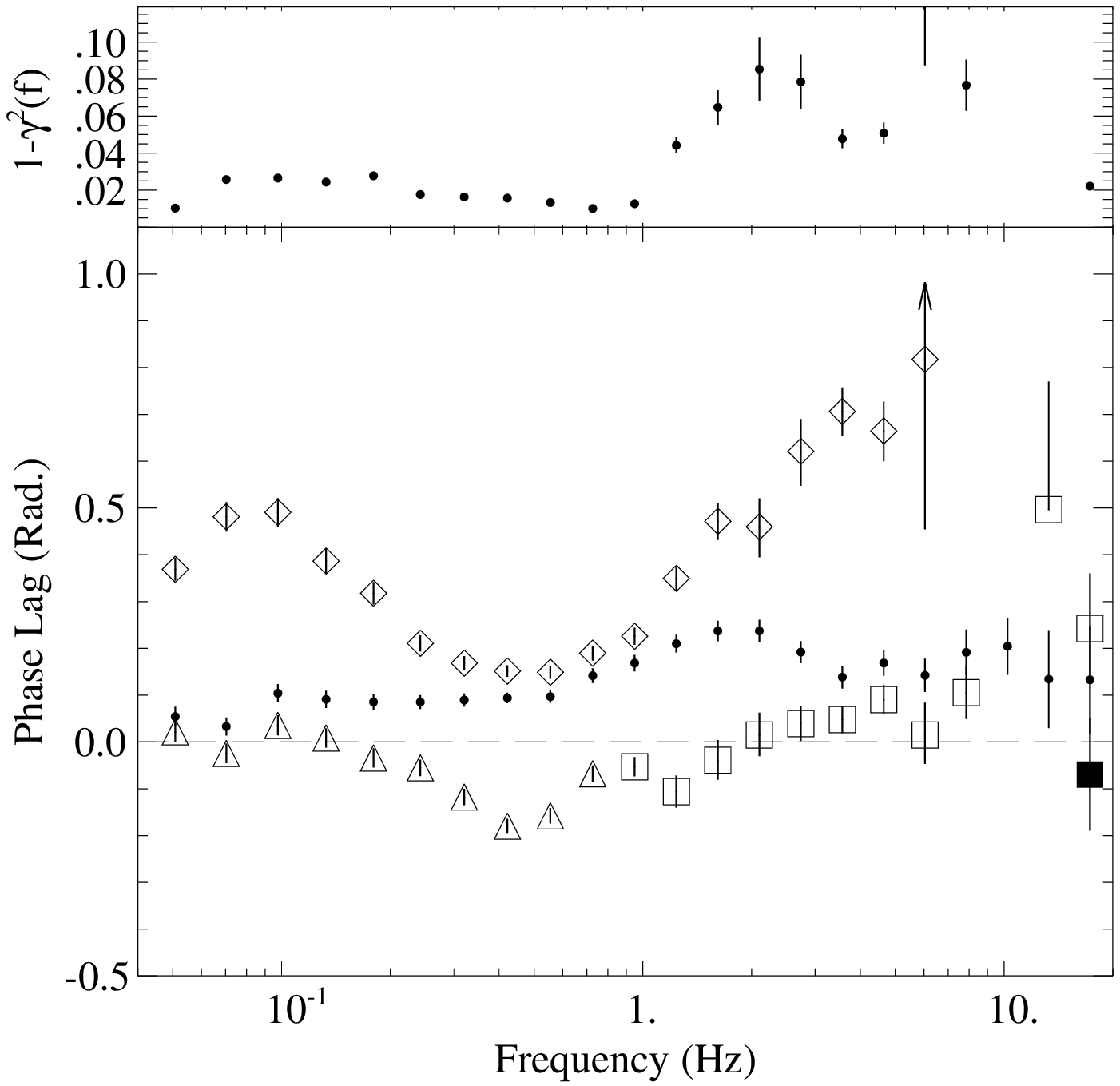} \caption{A composite
  power spectrum (presented as frequency $\times$ power) of GX~339$-$4
  in its hard state (left), along with the associated phase lags
  (right bottom) between hard and soft X-ray variability, and the
  coherence (presented as one minus coherence) of the phase lags
  (right top).  For the lag and coherence, filled circles are the
  data, while the clear triangles, diamonds, squares, and filled
  square are a possible decomposition of the lag into individual and
  independent components associated with each `broad feature' in the
  PSD (Nowak 2000).}\label{fig:psd}
\end{figure*} 

A potential example of such effects is shown in Fig.~\ref{fig:psd}.
These PSD, phase lags, and coherence (i.e., normalized amplitude of
the cross-correlation; \cite{vaughan:97a}) are composed from a set of
(very similar) observations of GX~339$-$4 \cite{nowak:00a}.  The PSD
is well modeled as a sum of broad features.  It is possible that
\emph{each} of these broad features has its own intrinsic phase/time
lag between hard and soft variability (a hypothesized decomposition is
shown in Fig.~\ref{fig:psd}), and that the net observed phase lag is
the sum of these components.  Furthermore, one would expect drops in
the coherence (i.e., the degree of linear correlation between soft and
hard variability) in regions where the independent PSD components
overlap \cite[see Fig.~\ref{fig:psd}, and][]{nowak:00a}.  (Again,
these thoughts are not new, and have been considered by Miyamoto and
collaborators for \textsl{Ginga} data of GBHC; \cite{miyamoto:88a}.)

Again, I believe the most fruitful avenue of research to pursue is to
start with a spectral decomposition and then work forward towards the
timing attributes, specifically, PSD, phase, and coherence.  A very
good example of this approach is the work by Poutanen and Gierlinski
\cite[][and these proceedings]{poutanen:03a}, who modeled the X-ray
spectra and variability of the pulsar SAX J1808.4$-$3658.  One of
their key results was to show that the measured phase could be related
to the individual spectral components.  This particular source,
however, had the advantage of being relatively bright and having a
strong variability feature (i.e., the pulse).  Thus there were good
statistics for performing such a spectral-temporal decomposition.  But
what about GBHC where most variability features are broad?

This latter case is difficult because obtaining variability phase
information requires extremely good statistics, and hence large
effective areas.  (Whereas there is an `optimal filter' to remove many
effects of Poisson noise from the PSD, there is no optimal filter to
minimize noise effects on phase measurements;
\cite{vanderklis:95a,vaughan:97a}.)  This seems to me a prime goal for
`spectral-temporal' studies that could be performed by a successor to
\rxte.  Also, given sufficient detector area \emph{and} the ability to
spectrally decompose rapid variability, we may yet consider performing
spectral-temporal studies in the time domain (as opposed to the
Fourier frequency domain).  It is worthwhile noting that, although
\conx\ is partly being designed with the goal of `reverberation
mapping' of the broad Fe line in AGN
\cite{reynolds:99a}, a similar study in GBHC not only requires a
faster time response, but also larger effective area.  The obtained
signal-to-noise in a spectrum integrated over a characteristic time
scale (viscous, thermal, or dynamical) is actually greater for AGN
compared to GBHC \cite[see the discussion of][and references
therein]{reynolds:03a}.  Again, this suggests a successor mission to
\rxte\ with its rapid timing capabilities, but substantially larger
effective area.

\section{Summary}

Over the past 30 years, we certainly have come a long way in our
understanding of the spectra (at all wavelength bands, not just X-ray)
of Galactic black hole binaries.  Although many of the components of
our theories and models have been in existence for a large fraction of
this history, it truly is the modern era, with four unique and
complementary X-ray satellites, working in cooperation with other
wavelength bands, wherein we are able to conduct careful tests of
these concepts.  Our capabilities will increase further (with the
launch of \swift\ and \astro, and the continuing missions of current
satellites) in the short-term future.  The farther future promises
more ambitious instruments, i.e., \conx, \xeus, and \maxim, which will
greatly enhance our knowledge of X-ray spectra.  

Still, if there is any cause for wistfulness, it will be the eventual
loss of the \rxte\ pointed instruments, \pca\ and \textsl{HEXTE}, and
the \rxte\ \textsl{All Sky Monitor}.  The flexibility of \rxte\ has
been crucial for multi-wavelength spectroscopy, and it has
reinvigorated the study of spectral-temporal correlations.  If there
is a `hole' to be filled in our, otherwise very exciting, future
studies of X-ray spectra, it is a new flexible satellite with rapid
timing capabilities and effective area larger than \rxte.


\begin{theacknowledgments}
  It is a pleasure to acknowledge stimulating conversations with Omer
  Blaes, Paolo Coppi, Sera Markoff, Jon Miller, and Christopher
  Reynolds. I would also like to thank my wife, Nirah Shomer, for
  suggesting the use of the 24 hour clock, and Jack Bauer for keep
  America safe.  I beg the indulgence of my boss for me not discussing
  the future of `high resolution X-ray spectroscopy', but that is the
  subject of a more thorough review by someone more knowledgeable than
  myself.  This work has been supported by NASA grants NAS8-01129 and
  GO3-4050B.
\end{theacknowledgments}



\end{document}